\documentclass[12pt]{article}
\usepackage{amssymb}
\usepackage{amsfonts}
\usepackage{graphicx}
\usepackage{psfrag}
\usepackage{epsfig}
\usepackage{bbm}
\usepackage{axodraw}

\newcommand{\inse}{{\Huge $\times$}}

\newcommand{\ini}{\begin{equation}}
\newcommand{\barray}{\begin{eqnarray}}
\newcommand{\fin}{\end{equation}}
\newcommand{\earray}{\end{eqnarray}}

\newcommand{\bsli}{\begin{slide}}
\newcommand{\esli}{\end{slide}}
\newcommand{\bcen}{\begin{center}}
\newcommand{\ecen}{\end{center}}

\newcommand{\bite}{\begin{itemize}}
\newcommand{\eite}{\end{itemize}}
\newcommand{\bmath}{\begin{displaymath}}

\newcommand{\emath}{\end{displaymath}}
\def\slash#1{\setbox0=\hbox{$#1$}#1\hskip-\wd0\hbox to\wd0{\hss\sl/\/\hss}}
\def\slashb#1{\setbox0=\hbox{$#1$}#1\hskip-\wd0\dimen0=5pt\advance
       \dimen0 by-\ht0\advance\dimen0 by\dp0\lower0.5\dimen0\hbox
         to\wd0{\hss\sl/\/\hss}}

\begin{document}

%\hyphenation{re-nor-ma-li-za-tion}

\begin{flushright}
%NYU-TH/02-09-11\\
hep-ph/0312242
\end{flushright}
\vspace{0.5cm}
\begin{center}
{\Large \bf Radiative Corrections and the Standard Model of Elementary
Particles\footnote{ To be published in the Annals of the National Academy of Exact, Physical, and Natural Sciences -- Buenos Aires, Argentina .}  }\\
%Talk given at the National Academy of Exact, Physical,
%and Natural Sciences -- Buenos Aires, Argentina on July 25, 2003.}  }\\
\vspace{0.7cm}
%{\Large\bf Eventual Comment.}\\
\vspace{0.5cm}
{\large 
% Giovanni Ossola\footnote{e-mail: go226@nyu.edu} and 
 Alberto Sirlin}
%\footnote{e-mail address: andrea.ferroglia@physics.nyu.edu}}\\

\vspace{1.5cm}
{\it Department of Physics, New York University,\\
4 Washington Place, New York, NY 10003, USA.} \\
\vspace{1.5cm}
\end{center}

\bigskip

\begin{center}
{\bf Abstract} 
\end{center}

This presentation includes an introductory discussion
of the unification of fundamental forces, properties of the 
elementary particles, Quantum Electrodynamics, the transition from
Quantum Electrodynamics and Weak Interactions to 
Electroweak Physics, Radiative Corrections in the Standard Model of
Particle Physics, and some critical unsolved problems.

\newpage

\section{Introduction}

The XIX century witnessed two great unifications in Physics. During the first 
decades of the century, major progress was achieved in understanding 
the interdependence of electric and magnetic phenomena by the work
of great physicists like Amp\`ere, Biot and Savart, Oersted, Faraday, and others.
This process culminated in Maxwell's theory, in which two separate areas of
Physics, Electricity and Magnetism, were unified in a single discipline, 
Classical Electromagnetism.

But Maxwell went beyond: studying the mathematical solutions of his equations, 
he reached the conclusion that they predict the existence of electromagnetic
waves that propagate in vacuum with the speed of light. Their presence
was soon confirmed experimentally by Hertz, and this discovery ushered
the era of modern telecommunications and technology. Maxwell then proposed 
that light is an electromagnetic wave in a restricted range of frequencies and, 
in a single stroke, achieved a second great unification, that of Optics
and Electromagnetism.

Maxwell's treatise on Classical Electromagnetism was published in 1874. 
Considered by many the most important theoretical physicist of the XIX century,
he died in 1879, at the early age of 47. The same year of 1879 witnessed the birth 
of his great successor, Albert Einstein. (Curiously enough, Newton was born in 1642,
the year of Galileo's death). In his Special Theory of Relativity, Einstein 
addressed some profound contradictions between Classical Electromagnetism
and Newtonian Mechanics. He left untouched Maxwell's theory, but altered 
Newtonian Mechanics and, in so doing, revolutionized our understanding of
space and time.

Approximately one hundred years later, a partial unification of three of the 
four fundamental forces of nature, the electromagnetic, weak and strong 
interactions, was achieved. This led to the emergence of the Standard Model
of Elementary Particles (1967-74), proposed originally by Weinberg, Salam, 
and Glashow, with very important contributions from other physicists.

\section{Brief Synopsis about Elementary Particles}

At present, physicists distinguish four fundamental forces in nature.
They manifest themselves as interactions between the fundamental particles.

1) Strong Interactions: responsible, for example, for the binding of
neutrons and protons in the atomic nuclei. They are a crucial factor
in nuclear physics.
2) Electromagnetic Interactions: responsible, for example, for the
binding of electrons and atomic nuclei to form atoms. They play 
a fundamental role in atomic physics, chemistry, and biology.
3) Weak Interactions: responsible for radioactive $\beta$ decays
(for instance, $n \to p + e^- + \overline{\nu}_e$), and for many decay
processes involving the elementary particles.
4) Gravitational Interactions: responsible, for example, for the dynamics
of the solar system. They are a crucial factor in the large scale structure 
of the universe.

These interactions are transmitted by certain fundamental particles that act
as mediators. For example, the electromagnetic interaction between two electrons
is described by the Feynman diagram

\begin{figure}[ht]
\begin{center}
\begin{picture}(300,100)(0,0)
\ArrowLine(100,50)(70,90) \Text(60,85)[]{$e'_1$}
\ArrowLine(70,10)(100,50) \Text(60,15)[]{$e_1$}
\ArrowLine(200,50)(230,90)\Text(240,85)[]{$e'_2$}
\ArrowLine(230,10)(200,50)\Text(240,15)[]{$e_2$}
\Photon(100,50)(200,50){4}{6}\Text(150,60)[]{$\gamma$}
\Vertex(100,50){1.5}
\Vertex(200,50){1.5}
\Text(150,5)[]{Fig.~1}
\end{picture}
\end{center}
%\caption{}
\end{figure} 

Here, an electron in the quantum state $e_1$ propagates in space-time
and emits a virtual photon $\gamma$ that also propagates in space-time
and interacts with the second electron. As a consequence of the interactions,
the two electrons change their quantum states from $e_1$ and $e_2$ to 
$e'_1$ and $e'_2$, respectively. The photon has zero mass and spin 1.

The weak interactions are mediated by three very massive intermediate bosons:
$W^{\pm}$, $Z^0$. Examples are 

\begin{figure}[ht]
\begin{center}
\begin{picture}(300,100)(0,0)
\ArrowLine(50,50)(30,90) \Text(20,85)[]{$p$}
\ArrowLine(30,10)(50,50) \Text(20,15)[]{$n$}
\ArrowLine(100,50)(115,90) \Text(120,85)[]{$e$}
\ArrowLine(130,80)(100,50) \Text(137,75)[]{$\overline{\nu}_e$}
\DashLine(50,50)(100,50){4} \Text(75,60)[]{$W$}
\Vertex(50,50){1.5}
\Vertex(100,50){1.5}
\ArrowLine(225,70)(180,90) \Text(175,85)[]{$\mu^-$}
\ArrowLine(270,90)(225,70) \Text(275,85)[]{$\mu^+$}
\ArrowLine(180,10)(225,30) \Text(175,20)[]{$e^-$}
\ArrowLine(225,30)(270,10) \Text(275,20)[]{$e^+$}
\DashLine(225,70)(225,30){4} \Text(238,50)[]{$Z^0$}
\Vertex(225,70){1.5}
\Vertex(225,30){1.5}
\Text(75,5)[]{Fig.~2a}
\Text(225,5)[]{Fig.~2b}
\end{picture}
\end{center}
%\caption{}
\end{figure}

The $W$ and $Z$ bosons are 
about 86 and 97 times more massive than the proton, respectively, and have spin 1.
Discovered in the decade 1980 to 1990, they were studied in great detail in major 
laboratories: CERN (Geneva, Switzerland), SLAC (Stanford, California), Fermilab
(ca. Chicago, Illinois). The present values of their masses are
$M_W = 80.426 \pm 0.034 \mbox{GeV}/c^2$, $M_Z = 91.1875 \pm 0.0021 
\mbox{GeV}/c^2$. ($\mbox{GeV}/c^2 \approx 1.8 \times 10^{-27} \mbox{Kg}$).
(In order to achieve the great precision in the measurement of $M_Z$, physicists
at CERN had to take into account the schedule of electric trains in the vicinity
of Geneva and the gravitational effects of the moon!). 

The present theory of strong interactions is called Quantum Chromodynamics.
The mediators are eight gluons. Like the photon, they have zero mass and spin 1.
They mediate the strong interactions between the quarks.

The fundamental matter fields are the leptons (which are not affected by 
the strong interactions) and the quarks (which are). They have spin $1/2$ and
appear in three generations, shown below.

\begin{displaymath}
\textrm{1st Generation}  \left\{ \begin{array}{lc} \nu_e & < 3\, \mbox{eV}/c^2 \\ 
                e & \approx 0.51\, \mbox{MeV}/c^2 \\ 
             u\, \textrm{(up)} & \textrm{(several)}\, \mbox{MeV}/c^2 \\ 
            d \,\textrm{(down)} &  \textrm{(several)}\, \mbox{MeV}/c^2
\end{array} \right.
\emath

\bmath
\textrm{2nd Generation} \quad \left\{ \begin{array}{lc} \nu_\mu & < 0.19\, \mbox{MeV}/c^2 \\ 
                \mu\, \textrm{(muon)} & \approx 106\, \mbox{MeV}/c^2 \\ 
             c\, \textrm{(charm)} & \approx 1.2\, \mbox{GeV}/c^2 \\ 
            s\, \textrm{(strange)} & \approx 120\,  \mbox{MeV}/c^2
\end{array} \right.
\end{displaymath}

\bmath
\textrm{3rd Generation} \quad \left\{ \begin{array}{lc} \nu_\tau & 
           < 18\, \mbox{MeV}/c^2 \\ 
                \tau\, \textrm{($\tau$-lepton)} & \approx 1.78\, \mbox{GeV}/c^2 \\ 
             t\, \textrm{(top)} & =(174.3 \pm 5.1)\, \mbox{GeV}/c^2 \\ 
            b\, \textrm{(bottom)} & \approx 4.3\,  \mbox{GeV}/c^2
\end{array} \right.
\end{displaymath}

\bmath
\textrm{Intermediate Bosons} \quad \left\{ \begin{array}{lc} \gamma &  0  \\ 
                W^\pm & =(80.426 \pm 0.034)\, \mbox{GeV}/c^2 \\ 
             Z^0 & =(91.1875 \pm 0.0021)\, \mbox{GeV}/c^2 \\ 
            g\, \textrm{(gluons)} & 0
\end{array} \right.
\end{displaymath}
 
\bmath
\textrm{Higgs Boson} \quad \left. \begin{array}{lc} H & > 115\, \mbox{GeV}/c^2            
\end{array} \right.
\emath

The charged leptons ($e$, $\mu$, $\tau$) have charge $-1$ in units of the proton 
charge, while the accompanying neutrinos ($\nu_e$, $\nu_\mu$, $\nu_\tau$) are neutral.
The quarks come in six flavors ($u$, $d$, $c$, $s$, $t$, $b$). The $u$, $c$, $t$
quarks have charge $2/3$ while  $d$, $s$, $b$ have charge $-1/3$.
Quarks and gluons are also endowed with another attribute called ``color''
that is associated with their strong interactions. On the right side we have
indicated the approximate masses ($\mbox{MeV}/c^2 \approx 1.8 \times 10^{-30}\, 
\mbox{kg}$) or upper and lower bounds.
Corresponding to each lepton or quark, there is an antiparticle with the 
same mass and opposite charge.

In the chart we have also included the intermediate bosons discussed before and the 
Higgs boson H of spin 0, a fundamental particle of the SM that so far has not
been discovered. Its interactions with the other particles play a crucial 
role in the generation of their masses.

Photons and neutrinos are very abundant. In each $\mbox{cm}^3$ of intergalactic space
there is an average of 412 photons and 112 neutrinos of each species.
The upper bounds on the neutrino masses are so small that, until recently,
the possibility existed that they may be massless. However,
in the last three years, very strong evidence has been found that they oscillate
among themselves. For example $\nu_e \to \nu_\mu$, $\nu_\mu \to \nu_\tau$,
which is only possible if they have mass. The study of neutrinos produced by the sun,
and of atmospheric neutrinos, has led to   
$| m_{\nu_\mu}^2 - m_{\nu_e}^2 | \approx 5 \times 10^{-5}\, \mbox{eV}^2$ and
$| m_{\nu_\tau}^2 - m_{\nu_\mu}^2 | \approx 3 \times 10^{-3}\, \mbox{eV}^2$,
respectively.

If we assume that $m_{\nu_\mu}^2  >>  m_{\nu_e}^2$,
we find $m_{\nu_\mu} \approx  7 \times 10^{-3}\, \mbox{eV}$,
much smaller than the present upper bound.

\section{Quantum Electrodynamics}

In the first three decades of the XX century two great revolutions 
took place in physics:

Special and General Relativity, developed by Einstein, and Quantum Mechanics,
associated with a large group of extraordinary physicists (Schr\"odinger, Heisenberg,
Bohr, Pauli, Born, Dirac, Einstein, Planck, \ldots).

The combination of Electromagnetism with  Quantum Mechanics and 
Special Relativity culminated in a very deep and successful theory
called Quantum Electrodynamics (QED) (Feynman, Schwinger, Tomonaga, Dyson, 
Bethe, \ldots).

As I mentioned before, in first approximation the interaction 
between two electrons can be described by Fig.1. Similarly, the interaction
of an electron with an external source, such as an atomic nucleus,
is represented by

\begin{figure}[ht]
\begin{center}
\begin{picture}(300,100)(0,0)
\ArrowLine(100,50)(70,90) \Text(60,85)[]{$e'_1$}
\ArrowLine(70,10)(100,50) \Text(60,15)[]{$e_1$}
%\ArrowLine(200,50)(230,90)\Text(240,85)[]{$e'_2$}
%\ArrowLine(230,10)(200,50)\Text(240,15)[]{$e_2$}
\Photon(100,50)(200,50){3}{8} \Text(150,60)[]{$\gamma$}
\Vertex(100,50){1.5}
%\Vertex(200,50){1.5}
\Text(200,50)[]{\inse}
%\Text(200,50)[]{\inse}
\Text(150,5)[]{Fig.~3}
\end{picture}
\end{center}
%\caption{}
\end{figure} 

Here, the \inse\, represents the atomic nucleus that emits a virtual
photon, which in turn interacts with the electron. Fig.~3 describes the scattering 
of the electron by its electromagnetic interaction with the nucleus.
QED leads to the conclusion that there are subtle corrections to this process
that can be evaluated systematically. An example is

\begin{figure}[ht]
\begin{center}
\begin{picture}(300,100)(0,0)
\ArrowLine(100,50)(79,78) 
\ArrowLine(79,78)(70,90) \Text(60,85)[]{$e'_1$}
\ArrowLine(70,10)(79,22) 
\ArrowLine(79,22)(100,50) \Text(60,15)[]{$e_1$}
%\ArrowLine(200,50)(230,90)\Text(240,85)[]{$e'_2$}
%\ArrowLine(230,10)(200,50)\Text(240,15)[]{$e_2$}
\Photon(100,50)(200,50){3}{8} \Text(150,60)[]{$\gamma$}
\Photon(79,78)(79,22){3}{5}  \Text(72,50)[]{$\gamma$}
%\Photon(100,50)(200,50){3}{8}
\Vertex(100,50){1.5} \Text(101,58)[]{$y$}
\Vertex(79,78){1.5}  \Text(84,82)[]{$z$}
\Vertex(79,22){1.5}  \Text(84,18)[]{$x$}
%\Vertex(200,50){1.5}
\Text(200,50)[]{\inse}
\Text(150,5)[]{Fig.~4}
%\Text(200,50)[]{\inse}
\end{picture}
\end{center}
%\caption{}
\end{figure}

Here, an electron propagates in space-time, emits a virtual photon at a 
space-time point x, at y absorbs the photon emitted by the nucleus , and 
at z absorbs the photon it has previously emitted.
Fig.~4 is called a Radiative Correction or a Quantum Correction
to the basic scattering process of Fig.~3.
In QED these corrections are evaluated as a series in powers of the fine
structure constant 
\bmath
\alpha = \frac{e^2}{\hbar c} = \frac{1}{137.03599877(40)} \, ,
\emath
where $e$ is the electron charge, $\hbar$ the Planck's constant,
$c$ the speed of light and (40) indicates the experimental error
that resides in the last two digits.
Each power of $\alpha$ corresponds to a loop in the Feynman diagram.
For instance, Fig.~4 contains one loop and leads to a correction
proportional to $\alpha$. There are other Feynman diagrams
in which, for example, the electron emits and absorbs $n$ virtual photons
leading to $n$ loops and a correction proportional to $\alpha^n$.

The calculation of the QED corrections leads to predictions of very high
precision.
For example, an electron possesses a magnetic moment 
${\vec m} = \frac{g e \hbar}{2 m_e c} {\vec S}$,
where ${\vec S}$ is its spin and $m_e$ its mass,
an attribute that governs its interaction with
magnetic fields. According to Dirac's theory of the electron,
which is a relativistic generalization of quantum mechanics,
$g = 2$.
Schwinger showed that the QED radiative corrections associated
with Fig.~4 alter this result leading to $a_e \equiv (g-2)/2 = \alpha/2 \pi$.
By now the corrections have been computed through ${\cal O}(\alpha^4)$ and
the experiments carried out with very high accuracy both for
the electron and the $\mu$-meson (muon).
The experimental and theoretical values are given below
\begin{eqnarray*}
a_{e^-}^{exp} & = & 1.1596521884(43) \times 10^{-3} \\
a_{e^+}^{exp} & = & 1.1596521879(43) \times 10^{-3} \\
a_{e}^{th} & = & 1.1596521640(160) \times 10^{-3} \\
& & \\
a_{\mu}^{exp} & = & 1.16592030(80) \times 10^{-3} \\
a_{\mu}^{th} & = & 1.16591693(78) \times 10^{-3} \\
a_{\mu}^{th} & = & 1.16591890(71) \times 10^{-3} 
\end{eqnarray*}
In the case of the muon, which is of great current interest, we have presented
two recent calculations in which subtle strong interaction effects have been
evaluated by different methods. At the present level of precision, these two 
calculations differ in the last digits, but it is expected that the origin of
this discrepancy will be understood better in the near future.

It is clear from the above numbers that QED is being tested with extraordinary
precision. For instance, in the electron case we are dealing with 11 significant
figures, with the error placed in the last two or three digits!

\section{Transition from QED and Weak Interactions to Electroweak Physics} 

The fact that the radiative corrections described in Section.~3 can be 
evaluated consistently is due to an important property of QED,
namely it is a renormalizable theory. In such theories, divergent
contributions to the mathematical expressions associated with Feynman
diagrams are eliminated as unobservable contributions to the fundamental
parameters. For instance, in QED the divergent parts in the evaluation
of the diagrams are absorbed as unobservable contributions to the mass
and charge of the electron. This process of removing the divergent 
contributions is called Renormalization.

The original theory of weak interactions, due to the great theoretical
and experimental physicist Enrico Fermi (1934), had remarkable
success in describing and relating a large number of phenomena
and experimental results. However, it was not renormalizable.
There were several attempts to create a renormalizable theory
of weak interactions, but in general they were not successful.
Finally, in the period 1967-1974 the Standard Model (SM) 
emerged with the aim of achieving
a unification of the weak, electromagnetic and strong interactions. 
Soon afterwords, the work of 't~Hooft, Veltman, Ben Lee, Zinn-Justin,
Becchi, Rouet, Stora, \ldots showed that the SM is renormalizable.
The original analysis of 't~Hooft and Veltman was constructed on the
basis of a technique, called Dimensional Regularization (DR), used to give
mathematical meaning to the divergent integrals associated with Feynman 
diagrams. Curiously enough, DR was invented, almost simultaneously,
in three different places: by Bollini and Giambiagi in Argentina,
't~Hooft and Veltman in the Netherlands, and Ashmore in Italy.

The renormalizability of the SM opened the possibility to study
the Radiative or Quantum Corrections in a consistent manner.
As the theory mixes the electromagnetic and weak contributions,
we call them Electroweak Corrections (EWC).

My first tentative steps in the study of the SM coincided with an
extremely fruitful visit to Argentina (January-August of 1972).
Soon after I arrived, I had a memorable conversation with Bollini
and Giambiagi, who explained to me the idea of DR.
Soon afterwords, we learned from Victor Alessandrini, who arrived
from Europe, that 't~Hooft and Veltman had also proposed DR in the
very important context of gauge theories.
During that visit I gave classes on the SM at the University of La Plata,
that were very well attended by people from La Plata and Buenos Aires,
and carried out my first research work in this area. 
With H.~Fanchiotti and H.~Girotti we discussed the cancellation of
ultraviolet divergencies in the unitary-gauge treatment of photon-photon
scattering, and with Bollini and Giambiagi the cancellation of ultraviolet
divergencies in natural relations of the SM.

My  principal area of interest since that time has been the study
of the EWC to important processes, with the aim to establish a close contact
between theory and high precision experiments.

The desideratum of these studies are 
\bite
\item[i)] To verify the SM at the level of its quantum corrections, attempting
to follow the great example of QED.
\item[ii)] To search for discrepancies between theory and experiment
or indications that may signal the presence of new physics beyond the SM.
\eite

These are the fundamental objectives 
of what is now known as Precision Electroweak Physics.

\section{Electroweak Corrections in the Standard Model}

I will give some illustrative examples.

\subsection{Unitarity of the Cabibbo-Kobayashi-Maskawa (CKM) matrix}

The interactions of the $W^\pm$ intermediate bosons with quarks
are governed by a matrix
\bmath
V = \left[ \begin{array}{ccc} 
V_{ud} & V_{us} & V_{ub} \\
V_{cd} & V_{cs} & V_{cb}  \\
V_{td} & V_{ts} & V_{tb} \end{array} \right]\, ,
\emath
where, for example, $V_{ud}$ refers to the coupling with the $u$, $d$ quarks.
According to the theory, an important property is that $V$ is a unitary matrix,
which implies $|V_{ud}|^2+|V_{us}|^2+|V_{ub}|^2=1 $ with analogous equalities 
for the other rows and columns of $V$. $|V_{ud}|$ is determined
by comparing the lifetimes of the muon and $\beta$ decays, which are known with
great experimental precision. On the theoretical side one needs the EWC to both 
processes. This problem had already been studied in the Fermi V-A theory
that preceded the SM (T.~Kinoshita, A.~Sirlin, S.M.~Berman (1958-59)).
At that time we found a great theoretical difficulty: while the EWC
to the muon lifetime were finite, those involving $\beta$ decay were
divergent. This was related to the fact that the Fermi theory  
of weak interactions is not renormalizable.

When the renormalizability of the SM was recognized in the early seventies,
I thought it was urgent to re-examine the problem in the light of the new theory.
In 1974 I obtained the answer in a simplified version of the SM, ignoring
the effect of the strong interactions (SI).
During 1974-78 I extended the analysis to the full-fledged SM, including
the effect of the SI. The EWC are now finite (as expected in a renormalizable 
theory) and of sizable magnitude.
They are dominated by a large logarithmic contribution $(3 \alpha / 2 \pi ) 
\ln{(M_Z / 2 E_m)} \sim 3.4\ \%$, where $E_m$ is the maximum energy of the
electron (or positron) in $\beta$ decay. It turns out that such large 
contribution was indeed required phenomenologically to satisfy the unitarity
of the $V$ matrix.
For me, that was the ``smoking gun'' of the SM at the level of the 
quantum corrections!

\subsection{Prediction of the $M_W$ and the $M_Z$ masses} \label{sb}

In 1979-81, William Marciano (who was a former student of mine) and I
thought that experimental physicists would attempt to discover the $W^\pm$
and $Z^0$ intermediate bosons, predicted by the SM, and measure their masses
$M_W$ and $M_Z$. It seemed a good idea to study, at the level of the EWC, the
relations between $M_W$, $M_Z$ and $G_F$, $\alpha$, as well as other
fundamental parameters of the theory such as the fermion masses, generically
denoted by $M_f$, and the mass $M_H$ of the Higgs boson.
Here $G_F = 1.16637(1) \times 10^{-5}\ \mbox{GeV}^{-2}$ is a very important 
constant that measures the magnitude of the weak interactions in the Fermi 
theory. We reached the conclusion that in order to predict $M_W$ and $M_Z$
it would be necessary to evaluate the EWC to a number of different processes
mediated by the $W^\pm$ and $Z^0$. As this seemed to be a difficult task
and the theoretical formulations at that time were very complicated, I 
thought the first step should be the development of a simple method to renormalize
the Electroweak Sector of the SM (A.~Sirlin, Physical Review D22, 971 (1980)).
This approach, with subsequent important contributions from other physicists,
is presently called the ``on-shell scheme of renormalization''.
Applying this scheme to muon decay, one finds
\bmath
\sin^2\theta_W \cos^2\theta_W = \frac{A^2}{M_Z^2 (1 - \Delta r)}\, ,
\emath
where $A^2 = \pi \alpha / (\sqrt{2} G_F)$, $\sin^2\theta_W = 1 - M_W^2/M_Z^2$
and $\Delta r$ is the corresponding EWC.
It has a complex structure and depends on several important parameters, 
in particular $M_H$, $M_W$, $M_Z$ and the top quark mass $M_t$.
In combination with experiments involving neutrino collisions with atomic nuclei
at high energies, the previous relations permitted to gradually improve
the predictions of $M_W$ and $M_Z$.

Around 1989, the great accelerators LEP at CERN and SLC at SLAC started operations
and Fermilab began the precision measurements of $M_W$. As mentioned in Sect.~1,
LEP soon determined $M_Z$ with great accuracy, and this led to a change in 
strategy: $\alpha$, $G_F$ and $M_Z$ were adopted as the basic input parameters,
and a major effort was done to study the $Z^0$ resonance in processes of the type
of Fig.~2b. In particular, several on resonance asymmetries and widths were 
measured with precision.
At present we know $\alpha$, $G_F$ and $M_Z$ with uncertainties of 
$\delta \alpha = \pm 0.0037\ \mbox{ppm}$, 
$\delta G_F = \pm 9\ \mbox{ppm}$,
$\delta M_Z = \pm 23\ \mbox{ppm}$, where ppm is an abbreviation for
``parts per million''.

In general, the experimental precision of the other observables is of the order of
$0.1 \%$ and this makes necessary to include the EWC in the theoretical predictions.

Thus, theorists working in this area were lucky: experimental physics in the 
great accelerators moved in the direction of high precision, where the
study of the EWC is particularly important!

\subsection{$M_t$ Prediction}

A very interesting example of the successful interplay between
theory and experiment was the $M_t$ prediction and its subsequent
measurement. Before 1995, the top quark could not be produced directly,
but it was possible to estimate its mass because of its contributions to 
the EWC. In Nov. 1994, a global analysis of the comparison between the SM
and the experiments led to the indirect determination
\bmath
M_t = 178 \pm 11^{+18}_{-19}\, \mbox{GeV}/c^2 \, ,
\emath
where the central value corresponds to $M_H = 300\, \mbox{GeV}/c^2$, the first error
is experimental and the second shift assumes 
$M_H = 65\, \mbox{GeV}/c^2 (-19\ \mbox{GeV}/c^2)$
or $M_H = 10^3\, \mbox{GeV}/c^2  (+18\ \mbox{GeV}/c^2)$.

Finally, with increasing energy and luminosity the top quark was produced at 
Fermilab and its mass measured. Its present value is
\bmath
M_t = 174.3 \pm 5.1\, \mbox{GeV}/c^2 \, ,
\emath 
very close to the prediction!

The possibility of this successful prediction is due to the fact that 
$\Delta r$ and other important EWC depend quadratically on $M_t$,
i.e. they contain contributions proportional to $M_t^2$ and are,
therefore, sensitive functions of $M_t$.

\subsection{The Higgs Boson}

This is a fundamental particle in the SM that, as mentioned in Section~2,
so far has not been found. Its interactions provide the mass of all
the other particles: intermediate vector bosons, leptons and quarks.
The direct search indicates that its mass $M_H \ge 114.4\, \mbox{GeV}/c^2$
at the $95\ \%$ confidence level.
With $M_t$ known experimentally, an important problem is the estimation of
$M_H$ by studying its effect on the EWC. This is much more difficult than
the $M_t$ prediction because the EWC depend only logarithmically on $M_H$,
and are therefore much less sensitive to this basic parameter.
One of the most important factors in constraining $M_H$ is the EWC $\Delta r$
mentioned in Subsection~\ref{sb}. The comparison between the theory and the 
current measurements of the various observables 
leads to the conclusion that $M_H < 211\, \mbox{GeV}/c^2$
at  the $95\ \%$ confidence level. We have therefore a limited band 
$114.4\, \mbox{GeV}/c^2 \lesssim  M_H \lesssim 211\, \mbox{GeV}/c^2$,
that will be explored at Fermilab and the new accelerator LHC (large
hadron collider) under construction at CERN.

\subsection{Supersymmetry}

One of the most interesting theoretical possibilities that involves new physics
beyond the SM is Supersymmetry (SUSY). It is a theory that, among many other
features, predicts that every boson (particle with integer spin) has a fermion
partner (particle of half-integer spin), and viceversa. In its simplest form
SUSY leads to an extension of the SM called MSSM (minimal supersymmetric
model). One of its most important predictions is that there are five Higgs
bosons and that the lightest one satisfies $M_H \lesssim 130\, \mbox{GeV}/c^2$.
The EWC have been also studied in great detail in the MSSM framework and play a
crucial role in the derivation of the $M_H$ upper bound.

On the other hand, supersymmetric partners of the usual elementary particles
have not been discovered thus far.
 
\subsection{Grand Unification}

The SM is invariant under certain mathematical transformations, called
gauge transformations, which are associated with a symmetry group
$SU(2)\times U(1) \times SU(3)$. The first two factor groups describe the 
symmetry properties of the Electroweak Sector of the theory, while the
third involves the Strong Interactions (QCD).
The three factor groups are characterized by parameters $g(\mu)$,
$g'(\mu)$, $g_s(\mu)$ that determine the magnitude of the interactions at the 
energy scale $\mu$ of the phenomena under consideration.
An attractive idea is the possibility that the three parameters are unified
at a high $\mu$ scale, where the symmetry is described by a single
factor group, such as $SU(5)$ or $SO(10)$.
This possibility is called Grand Unification. It has been shown that the three
lines defined by $g(\mu)$, $g'(\mu)$ and $g_s(\mu)$  as functions of $\mu$,
in fact intersect at $\mu \simeq 10^{16}\, \mbox{GeV}/c^2$ in the presence of 
SUSY, but not in its absence. This is one of the reasons that make the MSSM
theoretically attractive. In this analysis of Grand Unification the EWC
play also a very important role because they permit to obtain accurate values 
for $g(\mu)$, $g'(\mu)$ and $g_s(\mu)$ at the energy scale of the current 
experiments, which are important inputs in the calculations, and also 
govern the evolution of these parameters as functions of $\mu$. 

A major prediction of Grand Unified Theories, which has not been verified
yet, is that the proton is unstable, albeit with an extremely long lifetime.

\section{Conclusions}
\bite
\item[i)] The SM is a theory that describes with high precision a multitude of
phenomena from the atomic energy scale ($\simeq 10\, \mbox{eV}$) up to
$\simeq 100\, \mbox{GeV}$ (ten orders of magnitude!)
\item[ii)] It is a renormalizable theory so that its quantum corrections 
can be evaluated systematically, and this permits to compare the 
theoretical predictions with high precision experiments.
\item[iii)] This comparison has generally been very successful in
demonstrating\\ 
a) that the SM is correct at the $0.1\, \%$ level (assuming that the Higgs boson
will be found at a consistent mass scale), verifying the principle of gauge 
invariance, the symmetry group $SU(2)\times U(1) \times SU(3)$ and the
representations of this group assigned to the various particles.\\
b) that the EWC and QCD corrections are essentially correct,
verifying the validity of renormalizable gauge theories.
\item[iv)] The comparison has also permitted \\
c) to determine important parameters such as $\sin^2\theta_W$, to predict
$M_t$ and estimate $M_H$.\\
d) to sharply restrict possible new physics beyond the SM to be of a type
in which heavy new particles decouple at energies much lower than their
masses, such as supersymmetry.
\eite
 
A major unsolved problem is the unification of gravity with the other
three forces of nature and, more generally, the harmonious combination
of the two great revolutionary theories of the XX century, namely general
relativity and quantum theory.
Many theorists believe that string theory, in which elementary particles are
regarded as excitations of fundamental strings, rather than point structures,
offers the most promising paradigm to achieve these critical aims.

\section*{Acknowledgment}

This work was supported in part by NSF Grant PHY-0245068.

\end{document}